\pdfoutput=1

\documentclass[twocolumn]{autart}    

\usepackage{graphicx}          
\usepackage{amsmath}
\usepackage{amssymb}
\usepackage{fancyref}
\usepackage[dvipsnames]{xcolor}
\usepackage{hyperref}
\hypersetup{colorlinks=true, linkcolor=black, citecolor=black}
\usepackage{amsbsy}
\usepackage{bm}
\usepackage{caption}
\usepackage{emptypage}
\usepackage{multirow}
\usepackage{subcaption}
\usepackage{blindtext}
\usepackage{cases}
\usepackage{mathtools}
\usepackage{textcomp}
\usepackage{colortbl}
\usepackage{threeparttable}
\usepackage{booktabs}

\newcommand{\E}{\mathbb{E}}
\newcommand{\hF}{\hat{\mathcal{F}}}
\newcommand{\F}{\mathcal{F}}
\newtheorem{proposition}{Proposition}
\newtheorem{theorem}{Theorem}
\newtheorem{definition}{Definition}
\newtheorem{assumption}{Assumption}

\newcommand{\trace}{\mbox{Tr}}
\allowdisplaybreaks
\begin{document}
	
	\begin{frontmatter}

\title{Data-driven optimal control with a relaxed linear program\thanksref{footnoteinfo}} 

\thanks[footnoteinfo]{Research supported by the European Research Council under the H2020 Advanced Grant no. 787845 (OCAL). This paper was not presented at any conference. Corresponding author A. Martinelli.}

\author{Andrea Martinelli}\ead{andremar@ethz.ch},    
\author{Matilde Gargiani}\ead{gmatilde@ethz.ch},               
\author{John Lygeros}\ead{lygeros@ethz.ch}  

\address{Automatic Control Laboratory, ETH Zurich, Physikstrasse 3, 8092 Zurich, Switzerland}            
          
\begin{keyword}                           
  Approximate dynamic programming, Optimal control, Data-driven control, Linear programming, Stochastic systems                                       
\end{keyword}                             

\begin{abstract}                          
The linear programming (LP) approach has a long history in the theory of approximate dynamic programming. When it comes to computation, however, the LP approach often suffers from poor scalability. In this work, we introduce a relaxed version of the Bellman operator for $q$-functions and prove that it is still a monotone contraction mapping with a unique fixed point. In the spirit of the LP approach, we exploit the new operator to build a relaxed linear program (RLP). Compared to the standard LP formulation, our RLP has only one family of constraints and half the decision variables, making it more scalable and computationally efficient. For deterministic systems, the RLP trivially returns the correct $q$-function. For stochastic linear systems in continuous spaces, the solution to the RLP preserves the minimizer of the optimal $q$-function, hence retrieves the optimal policy. Theoretical results are backed up in simulation where we solve sampled versions of the LPs with data collected by interacting with the environment. For general nonlinear systems, we observe that the RLP again tends to preserve the minimizers of the solution to the LP, though the relative performance is influenced by the specific geometry of the problem.
\end{abstract}

\end{frontmatter}

\section{Introduction}

The term \textit{optimal control} came into use in the 1950s to describe the problem of designing a controller to minimize a measure of a dynamical system's behavior over time \cite{SuttonRLanIntroduction}. The problem formulation is widely applicable and it arises in many disciplines, from robotics to bioengineering to finance \cite{BoborowRobotics,KuoBioengineering,BERTSIMASFinance}, to name a few. In the same years, R. Bellman developed a method, based on the principle of optimality, that uses the concept of value function to define a functional equation -- the Bellman equation \cite{BellmanDP}. The class of methods for solving optimal control problems by solving this equation came to be known as \textit{dynamic programming} (DP). For computing the value function, DP methods typically rely on three fundamental approaches: value iteration, policy iteration and linear programming (LP) \cite{BertsekasVol2}. The mathematical foundations of these approaches lie in the monotonicity and contractivity properties of a functional operator, implicitly defined in the Bellman equation \cite{BertsekasAbstractDP}. DP methods suffer from what Bellman called the \textit{curse of dimensionality}, meaning that the computational requirements grow exponentially with the number of state and input variables. Several approximation methods have been developed to mitigate this curse of dimensionality, collectively known as \textit{approximate} dynamic programming (ADP) \cite{BertsekasVol2,PowellADP}. 

In this context, our work focuses on the LP approach to ADP. The LP approach, introduced by \cite{ManneLP} in the 60s, exploits the properties of the Bellman operator to build linear programs whose solution coincides with the optimal value function. In the context of ADP with continuous state or action spaces, the exact infinite dimensional LPs are approximated by tractable finite dimensional ones \cite{SchweitzerPolynomApproxinMDP,LasserreDTMCP,deFariasLPapproach,EsfahaniFromInftoFinitePrograms}.

In many real-world applications one has to operate a system without the knowledge of its dynamical equation. Learning how to optimally regulate a system is possible by letting it interact with the environment and collecting costs (or rewards) over time, an approach classically known as \textit{reinforcement learning} (RL) \cite{SuttonRLanIntroduction,SuttonTemporalDiff,BUSONIU2018}. To extend RL methods in the LP approach framework, one can reformulate the Bellman equation in terms of the $q$-function \cite{WatkinsQLearning}, and set up a new class of LPs based on the Bellman operator for $q$-functions \cite{CogillDecentralizedADP,PaulADP}. The main advantage of the $q$-function formulation is that one can extract a policy directly from $q$, without knowledge of the system dynamics or stage cost. The problem of learning the optimal $q$-function from data with the LP approach has been recently addressed for both deterministic \cite{GoranADP,AlexandrosIFAC20,MayneConvexQLearning} and stochastic \cite{SutterCDC2017} systems.

The contributions of this work can be summarized as:
\begin{itemize}
	\item Introduction of a relaxed Bellman operator which preserves monotonicity and contractivity properties;
	\item Construction of an associated infinite dimensional relaxed linear program (RLP) which is more scalable and efficient. In the case of deterministic systems, the RLP trivially returns the correct $q$-function;
	\item Fixed point characterization in the case of stochastic linear systems with quadratic costs: the RLP preserves the minimizer of the optimal $q$-function;
	\item Data-driven implementation and analysis of the LP and RLP in both linear and nonlinear settings. 
\end{itemize}
In the next section, we introduce the necessary background on the LP approach to ADP.


\section{The linear programming formulation}

\subsection{Mathematical preliminaries}
Let $\mathbf{Y}$ be a finite dimensional vector space. We introduce a weight function $r : \mathbf{Y}\rightarrow\mathbb{R}$ such that $r(y) > 0 \; \forall y\in\mathbf{Y}$, and denote with $\mathcal{S}(\mathbf{Y})$ the vector space of all real-valued measurable functions $g : \mathbf{Y}\rightarrow\mathbb{R}$ that have a finite weighted sup-norm \cite[\S 2.1]{BertsekasAbstractDP}
\begin{equation}\label{finite sup-norm}
	|| g(x) ||_{\infty,r} = \sup_{y\in\mathbf{Y}}\frac{|g(y)|}{r(y)}<\infty.
\end{equation}
The following definitions and theorem can be found in \cite[\S 1]{RockafellarOperatorTheory} and \cite[Def. 5.1-1 and Thm. 5.1-2]{KreyszigFunctionalAnalysis}, respectively.
\begin{definition}\label{Def Monotonocity}
	A map $\mathcal{J} : \mathcal{S}(\mathbf{Y}) \rightarrow \mathcal{S}(\mathbf{Y})$ is monotone iff
	\begin{equation*}
		\langle\mathcal{J}g_1 - \mathcal{J}g_2, \, g_1-g_2\rangle \ge 0 \quad \forall g_1,g_2 \in \mathcal{S}(\mathbf{Y}).
	\end{equation*}
\end{definition} 
Next, consider a metric $d$ on the space $\mathcal{S}(\mathbf{Y})$, making $(\mathcal{S}(\mathbf{Y}),d)$ a metric space. 
\begin{definition}\label{Def Contraction}
	A map $\mathcal{J} : \mathcal{S}(\mathbf{Y}) \rightarrow \mathcal{S}(\mathbf{Y})$ is a contraction with respect to the metric $d$ iff there exists a constant $k \in [0,1)$ such that 
	\begin{equation*}
		d(\mathcal{J}g_1,\mathcal{J}g_2) \le k d(g_1,g_2) \quad \forall g_1,g_2 \in \mathcal{S}(\mathbf{Y}).
	\end{equation*}
\end{definition}

\begin{theorem}\label{Banach Theorem}
	Let ($\mathcal{S}(\mathbf{Y}),d$) be a complete metric space with a contraction $\mathcal{J}$. Then, $\mathcal{J}$ has a unique fixed point, i.e. there exists a unique $\bar{y}\in \mathbf{Y}$ such that $\bar{y}=\mathcal{J}\bar{y}$.
\end{theorem}

\subsection{Stochastic optimal control}

Consider a discrete-time stochastic dynamical system
\begin{equation}\label{nonlinear dynamical system}
	x_{k+1} = f(x_k,u_k,\xi_k),
\end{equation}
with (possibly infinite) state and action spaces $x_k \in \mathbf{X} \subseteq \mathbb{R}^{n_{x}}$ and $u_k \in \mathbf{U} \subseteq \mathbb{R}^{n_{u}}$. Here, $\xi_k \in \mathbf{\Xi} \subseteq \mathbb{R}^{n_{\xi}}$ denotes the realizations of independent identically distributed (i.i.d.) random variables, and $f : \mathbf{K} \times \mathbf{\Xi} \rightarrow \mathbf{X}$, with $\mathbf{K} = \mathbf{X} \times \mathbf{U}$, is the map encoding the dynamics.	
We consider \textit{stationary feedback policies}, given by functions $\pi : \mathbf{X} \rightarrow \mathbf{U}$; for more general classes of policies, see \cite{LasserreDTMCP}. A nonnegative cost is associated to each state-action pair through the \textit{stage cost} function $\ell : \mathbf{K} \rightarrow \mathbf{R}_{+}$. We introduce a \textit{discount factor} $\gamma \in (0,1)$ and consider the infinite-horizon cost associated to policy $\pi$
\begin{equation}
	v_{\pi}(x) = \E_{\xi}\left[  \sum_{k=0}^{\infty}\gamma^{k}\ell(x_k,\pi(x_k)) \; \bigg| \; x_0 = x \right].
\end{equation}
The objective of the optimal control problem is to find an optimal policy $\pi^{\ast}$ such that $v_{\pi^{*}}(x) = \inf_{\pi} v_{\pi}(x) = v^{*}(x)$, where $v^*$ is known as the optimal \textit{value function}. Throughout the paper, we work under \cite[Assump. 4.2.1 and 4.2.2]{LasserreDTMCP} to ensure that $v^{*} \in \mathcal{S}(\mathbf{X})$, $\pi^{*}$ is measurable and the infimum of $v_{\pi}$ is attained. 

\subsection{Contractions and linear programs}\label{Section 2}

The optimal policy $\pi^{*}$ is generally difficult to compute since, amongst other issues, it involves the minimization of an infinite sum of costs. However, the value function associated to $\pi$ can be recursively defined as \cite{BellmanDP}
\begin{equation}\label{Bellman equation}
	v_{\pi}(x) = \ell(x,\pi(x)) + \gamma \E_{\xi}\big[v_{\pi}(f(x,\pi(x),\xi))\big],
\end{equation}
for all $x \in \mathbf{X}$. From now on, whenever possible we will denote $f(x,\pi(x),\xi)$ as $x^{+}_{\pi}$. Equation \eqref{Bellman equation}, of course, also holds for an optimal policy
\begin{align}
	v^{*}(x) & = \ell(x,\pi^{*}(x)) + \gamma \E_{\xi}\big[v_{\pi^{*}}(x^{+}_{\pi^{*}})\big] \notag \\
	& = \inf_{u\in \mathbf{U}} \bigl\{ \ell(x,u) + \gamma \E_{\xi}\big[v^{*}(x^{+}_{u})\big] \bigr\} \notag \\
	& = \mathcal{T}v^{*}(x) \quad \forall x \in \mathbf{X}.
\end{align}
The operator $\mathcal{T}$ is the \textit{Bellman operator}, it maps from $\mathcal{S}(\mathbf{X})$ to itself and can be shown to possess the two fundamental properties of \textit{monotonicity} (Definition \ref{Def Monotonocity}) and \textit{$\gamma$-contractivity} with respect to the sup-norm (Definition \ref{Def Contraction}) \cite{BertsekasAbstractDP,DenardoContractionMaps}.
Thanks to Theorem \ref{Banach Theorem} we are then guaranteed that $\mathcal{T}$ has a unique fixed point $\bar{v}$ given by 
\begin{equation}
	\bar{v} = \mathcal{T}\bar{v} = v^{*} = \lim_{n \rightarrow \infty}\mathcal{T}^{n}v \quad \forall v \in \mathcal{S}(\mathbf{X}).
\end{equation}
By exploiting the monotonicity and contractivity properties, we observe that the Bellman inequality $v \le \mathcal{T}v$ implies $v \le v^{*}$. It is then natural to look for the greatest $v \in \mathcal{S}(\mathbf{X})$ that satisfies $v \le \mathcal{T}v$:
\begin{equation}\label{NPvaluefunction}
	\begin{aligned}
		\sup_{v\in\mathcal{S}(\mathbf{X})} & \int_{\mathbf{X}} v(x)c(dx) \\
		\mbox{s.t.} \;\; & v(x) \le \mathcal{T}v(x) \quad \forall x \in \mathbf{X},
	\end{aligned}
\end{equation}
where $c$ is a non-negative measure that assigns positive mass to all open subsets of $\mathbf{X}$. Notice that $\mathcal{T}$ is a nonlinear operator. However, it is possible to reformulate \eqref{NPvaluefunction} as an equivalent linear program \cite{deFariasLPapproach} by dropping the infimum in $\mathcal{T}$ and substituting the nonlinear constraint set with the following linear one
\[ v(x) \le \mathcal{T}_Lv(x,u) = \ell(x,u) + \gamma \E_{\xi}\big[v(x^{+}_{u})\big] \; \forall (x,u) \in \mathbf{K}.  \] 
The obtained formulation is in general an infinite dimensional linear program, and it is not solvable due to several sources of intractability, which are collectively known as \textit{curse of dimensionality}, see e.g. \cite{PaulADP} and \cite{WangIteratedBellInequalities}. 

If one is nonetheless able to obtain $v^*$, they can in principle compute the corresponding policy by 
\begin{equation}
	\pi^*(x) = \arg \min_{u\in\mathbf{U}} \big\{ \ell(x,u) + \gamma \E_{\xi}\big[v^*(x^{+}_{u})\big] \big\}.
\end{equation}
However, if the dynamics $f$ or the stage cost $\ell$ are not known, this calculation is also impossible. This difficulty can be addressed by introducing the $q$-function \cite{WatkinsQLearning} associated to a policy $\pi$ as
\begin{align}
	q_{\pi}(x,u) & = \ell(x,u) + \gamma \E_{\xi}\big[v_{\pi}(x^{+}_{u})\big] \notag \\
	& = \ell(x,u) + \gamma \E_{\xi}\big[q_{\pi}(x^{+}_{u},\pi(x^{+}_{u}))\big],
\end{align}
for all $(x,u) \in \mathbf{K}$. This can be interpreted as the cost of applying control input $u$ at state $x$, and following policy $\pi$ thereafter. The optimal $q$-function is expressed by
\begin{align}\label{Bellman operator q functions}
	q^{*}(x,u) & = \ell(x,u) + \gamma \E_{\xi}\left[\inf_{w\in \mathbf{U}}q_{\pi^{*}}(x^{+}_{\pi^{*}},w)\right] \notag \\
	& = \mathcal{F}q^{*}(x,u) \quad \forall(x,u) \in \mathbf{K}.
\end{align}
The link between $v^*$ and $q^*$ is then given by 
\begin{equation}
	v^*(x) = \inf_{u\in \mathbf{U}}q^*(x,u).
\end{equation}
The advantage of the $q$-function reformulation is that the policy extraction does not require knowledge of $f$ and $\ell$:
\begin{equation}
	\pi^*(x) = \arg \min_{u\in\mathbf{U}}q^*(x,u).
\end{equation}
Since the operator $\mathcal{F}$ shares the same monotonocity and contractivity properties of $\mathcal{T}$ \cite{BertsekasNDP}, we can write again a (nonlinear) exact program for the $q$-function
\begin{equation}\label{NonlinearProgramQfunctions}
	\begin{aligned}
		\sup_{q\in\mathcal{S}(\mathbf{K})} & \int_{\mathbf{K}} q(x,u)c(dx,du) \\
		\mbox{s.t.} \;\; & q(x,u) \le \mathcal{F}q(x,u) \quad \forall (x,u) \in \mathbf{K},
	\end{aligned}
\end{equation}
where $c$ takes the same role than in \eqref{NPvaluefunction}. Unlike \eqref{NPvaluefunction}, it is not straightforward to replace the nonlinear constraints in \eqref{NonlinearProgramQfunctions} with linear ones due to the nesting of the $\E$ and $\inf$ operators in \eqref{Bellman operator q functions}. A linear reformulation of \eqref{NonlinearProgramQfunctions} can be obtained, as shown in \cite{CogillDecentralizedADP,PaulADP}, by introducing additional decision variables
\begin{equation}\label{LPoldQfunctions}
	\begin{aligned}
		\sup_{\scriptsize
				v\in \mathcal{S}(\mathbf{X}), q\in\mathcal{S}(\mathbf{K})
		} & \int_{\mathbf{K}} q(x,u)c(dx,du) \\
		\mbox{s.t.} \;\; & q(x,u) \le \mathcal{T}_Lv(x,u) \quad \forall (x,u) \in \mathbf{K} \\
		& v(x) \le q(x,u) \quad \forall (x,u) \in \mathbf{K}.
	\end{aligned}
\end{equation}
Following \cite{GoranADP}, in the case of deterministic systems the lack of expectation makes possible to compute $q^*$ by solving the simpler LP
\begin{equation}\label{LPdeterministic}
	\begin{aligned}
		\sup_{q\in \mathcal{S}(\mathbf{K})} & \int_{\mathbf{K}} q(x,u)c(dx,du) \\
		\mbox{s.t.} \;\; & q(x,u) \le \ell(x,u) + \gamma q(f(x,u),w),
	\end{aligned}
\end{equation}
for all $(x,u,w) \in \mathbf{H} = \mathbf{K}\times\mathbf{U}$. Our aim is to derive a simplified version of \eqref{LPoldQfunctions} for stochastic systems, inspired by its deterministic counterpart \eqref{LPdeterministic}. That is, a program that involves $q$ only and a single family of constraints.

\section{The relaxed linear program}

We introduce a new functional operator.
\begin{definition}
	The relaxed Bellman operator is the map $\hF:\mathcal{S}(\mathbf{K})\rightarrow\mathcal{S}(\mathbf{K})$ given by
	\begin{equation}\label{relaxed operator}
		\hF q(x,u) = \ell(x,u) + \gamma \inf_{w} \E_{\xi}\big[q(x^{+}_{u},w)\big].
	\end{equation}
\end{definition}
Note that the operator \eqref{relaxed operator} retains the same structure as the standard Bellman operator \eqref{Bellman operator q functions}, but the expectation and infimum are exchanged. In the following, we show some fundamental properties of \eqref{relaxed operator}. 
\begin{proposition}\label{Prop monotonicity}
	The operator $\hF$ is a monotone contraction mapping with a unique fixed point in $\mathcal{S}(\mathbf{K})$.
\end{proposition}
\begin{pf}
	(i) \textit{Monotonicity}. Consider $q_1,q_2 \in \mathcal{S}(\mathbf{K})$ such that $q_1(x,w) \le q_2(x,w) \; \forall (x,w)$. Hence, 
	\begin{align*}
		 q_1(x^{+}_{u},w) & \le q_2(x^{+}_{u},w) \quad \forall (x,u,w,\xi), \\
		 \E_{\xi}\big[q_1(x^{+}_{u},w)\big] & \le \E_{\xi}\big[q_2(x^{+}_{u},w)\big] \quad \forall (x,u,w), \\
		  \inf_{w}\E_{\xi}\big[q_1(x^{+}_{u},w)\big] & \le \inf_{w}\E_{\xi}\big[q_2(x^{+}_{u},w)\big] \quad \forall (x,u),& \\
		  \hF q_1(x,w) & \le \hF q_2(x,w) \quad \forall (x,w).   	
	\end{align*}
	Therefore, $\langle\hF q_1 - \hF q_2,\, q_1 - q_2\rangle \ge 0 \;\, \forall q_1,q_2 \in \mathcal{S}(\mathbf{K})$, and the operator is monotone. \\	
	(ii) \textit{Contractivity}. Given $q_1,q_2 \in \mathcal{S}(\mathbf{K})$, we have that
	\begin{multline*}
		\left| \hF q_1(x,w) - \hF q_2(x,w) \right| = \\
		\begin{aligned}
			& = \gamma\left| \inf_{w} \E_{\xi}\left[q_1(x^{+}_{u},w)\right] - \inf_{w}  \E_{\xi}\left[q_2(x^{+}_{u},w)\right]  \right| \\
			& \le \gamma\sup_{w}\left| \E_{\xi}\left[q_1(x^{+}_{u},w)\right] - \E_{\xi}\left[q_2(x^{+}_{u},w)\right] \right| \\
			& = \gamma\sup_{w}\left| \E_{\xi}\left[q_1(x^{+}_{u},w) - q_2(x^{+}_{u},w)\right] \right| \\
			& \le \gamma\sup_{w}\E_{\xi} \left| q_1(x^{+}_{u},w) - q_2(x^{+}_{u},w) \right| \\
			& \le \gamma \sup_{x,w}\left| q_1(x,w) - q_2(x,w) \right|.
		\end{aligned}
	\end{multline*}
	Hence $||\hF q_1 - \hF q_2||_{\infty} \le \gamma ||q_1 - q_2||_{\infty} \; \forall q_1,q_2 \in \mathcal{S}(\mathbf{K})$,
	and $\hat{\mathcal{F}}$ is a $\gamma$-contraction with respect to the sup-norm. \\	
	(iii) \textit{Uniqueness of fixed point}. As the vector space $\mathcal{S}(\mathbf{K})$ is complete under the sup-norm \cite[\S B.2]{BertsekasAbstractDP}, the result follows from (ii) and Theorem \ref{Banach Theorem}. \qed
\end{pf}
\begin{proposition}\label{Prop difference}
	The fixed point of $\hF$ is a point-wise upper bound to the fixed point of $\mathcal{F}$.
\end{proposition}
\begin{pf}
	Since $\inf_{w}q(x^{+}_{u},w) \le q(x^{+}_{u},w) \; \forall (x,u,w,\xi)$,
	\begin{align}
		 \E_{\xi}\big[\inf_{w}q(x^{+}_{u},w)\big] & \le \E_{\xi}\big[q(x^{+}_{u},w)\big] \quad \forall (x,u,w), \notag \\
			\E_{\xi}\big[\inf_{w}q(x^{+}_{u},w)\big] & \le \inf_{w}\E_{\xi}\big[q(x^{+}_{u},w)\big] \quad \forall (x,u), \notag \\
		 \F q(x,w) & \le \hF q(x,w) \quad \forall q \in \mathcal{S}(\mathbf{K}), \label{inequality}
	\end{align}
	which implies $q^* \le \hat{q}$. \qed
\end{pf}
Now consider the (nonlinear) program
\begin{equation}\label{alternativeNPQfunctions}
	\begin{aligned}
		\sup_{q\in\mathcal{S}(\mathbf{K})} & \int_{\mathbf{K}} q(x,u)c(dx,du) \\
		\mbox{s.t.} \;\; & q(x,u) \le \hat{\mathcal{F}}q(x,u) \quad \forall (x,u) \in \mathbf{K}.
	\end{aligned}
\end{equation}
\begin{proposition}\label{feasibilitysetNP}
	If $q$ is feasible for \eqref{NonlinearProgramQfunctions}, then is feasible for \eqref{alternativeNPQfunctions}. Moreover, the unique optimal solution to \eqref{alternativeNPQfunctions} is the fixed point of $\hF$.
\end{proposition}
\begin{pf}
	According to inequality \eqref{inequality}, if $q$ is feasible for \eqref{NonlinearProgramQfunctions} then $q\le \mathcal{F}q \le \hF q \;\, \forall (x,u) \in \mathbf{K}$. The second statement follows from Proposition \ref{Prop monotonicity}. \qed
\end{pf}
On the same line of the linearizations in \eqref{NPvaluefunction} and \eqref{NonlinearProgramQfunctions}, one can replace the nonlinear constraints in \eqref{alternativeNPQfunctions} with linear ones and obtain the \textit{relaxed linear program} (RLP)
\begin{equation}\label{alternativeLPQfunctions}
	\begin{aligned}
		\sup_{q\in\mathcal{S}(\mathbf{K})} & \int_{\mathbf{K}} q(x,u)c(dx,du) \\
		\mbox{s.t.} \;\; & q(x,u) \le \hat{\mathcal{F}}_{L}q(x,u,w) \quad \forall (x,u,w) \in \mathbf{H},
	\end{aligned}
\end{equation}
where $\hF_{L} q(x,u,w) = \ell(x,u) + \gamma \mathbb{E}_{\xi}\left[q(x^{+}_{u},w)\right]$.
\begin{theorem}\label{feasibility set}
	If $(q,v)$ is feasible for \eqref{LPoldQfunctions}, then $q$ is feasible for \eqref{alternativeLPQfunctions}. Moreover, the unique optimal solution to the RLP \eqref{alternativeLPQfunctions} is the fixed point of $\hF$.
\end{theorem}
\begin{pf}
	If $(q,v)$ is a feasible pair for \eqref{LPoldQfunctions}, then $q \le \ell + \gamma \E_{\xi}v \le \ell + \gamma \E_{\xi}q = \hF_{L}q \;\, \forall (x,u,w) \in \mathbf{H}$. Furthermore, thanks to Proposition \ref{feasibilitysetNP}, we know that $\hat{q}$ is the unique optimal solution to \eqref{alternativeNPQfunctions}. As the RLP \eqref{alternativeLPQfunctions} is a relaxation of \eqref{alternativeNPQfunctions}, $\hat{q}$ is feasible for \eqref{alternativeLPQfunctions}. On the other hand, any feasible solution $q'$ to \eqref{alternativeLPQfunctions} satisfies $q \le \hat{\mathcal{F}}_{L}q$ for all $(x,u,w) \in \mathbf{H}$ and, in particular, for the $w$ that minimizes $\E_{\xi}q(x^{+}_{u},w)$. That is, $q'$ satisfies $q \le \hF q$ and therefore it is a lower bound to $\hat{q}$. As a consequence, $\hat{q}$ is the unique optimal solution to \eqref{alternativeLPQfunctions}. \qed
\end{pf}
Note that, according to Propositions \ref{feasibilitysetNP} and Theorem \ref{feasibility set}, the programs \eqref{alternativeNPQfunctions}-\eqref{alternativeLPQfunctions} are relaxations of \eqref{NonlinearProgramQfunctions}-\eqref{LPoldQfunctions}, respectively. Hence, the name of \textit{relaxed} Bellman operator for $\hF$. In contrast to \eqref{LPoldQfunctions}, however, the RLP requires $q$ only and involves only one family of constraints. According to Proposition \ref{Prop difference}, the optimal solution to the RLP is an upper bound to $q^*$. Note that for deterministic systems the RLP reduces to \eqref{LPdeterministic}, hence returns $q^*$ as its optimizer. This allows to simplify existing deterministic LP formulations such as the one in \cite{MayneConvexQLearning}.

\section{Fixed point analysis for linear systems}

Next, we turn our attention to the relation between the optimisers of \eqref{LPoldQfunctions} and \eqref{alternativeLPQfunctions} and the corresponding optimal policies. A special case of the infinite-horizon optimal control problem arises when the dynamics is linear 
\begin{equation}\label{linearmap}
	f(x,u,\xi) = Ax + Bu + \xi,
\end{equation}
with $\mathbf{X} = \mathbb{R}^{n_{x}}$, $\mathbf{U} = \mathbb{R}^{n_{u}}$, $\mathbf{\Xi} = \mathbb{R}^{n_{x}}$, $A \in \mathbb{R}^{n_{x}\times n_{x}}, B \in \mathbb{R}^{n_{u}\times n_{x}}$, and the cost function is quadratic
\begin{equation}\label{quadraticcost}
	\ell(x,u) = \begin{bmatrix} x \\ u \end{bmatrix}^{\top} L \begin{bmatrix} x \\ u \end{bmatrix} = \begin{bmatrix} x \\ u \end{bmatrix}^{\top} \begin{bmatrix}
			L_{xx} & L_{xu} \\ L_{xu}^{\top} & L_{uu}
	\end{bmatrix} \begin{bmatrix} x \\ u \end{bmatrix}. 
\end{equation}

\begin{assumption}\label{assumtion stabilizability}
	The pair $(\gamma^{\frac{1}{2}}A,\gamma^{\frac{1}{2}}B)$ is stabilizable, $\xi$ is i.i.d. with zero mean and covariance matrix $\Sigma$, $L$ is positive semi-definite and $L_{uu}$ is positive definite.
\end{assumption}
Note that in this case quadratic functions belong to $\mathcal{S}(\mathbf{K})$ for an appropriate quadratic weight function $r(x)$. One can then show that the optimal $q$-function for such a problem is \cite{DavisStochasticControl}
\begin{gather}\label{qstar}
	q^{*}(x,u) = \begin{bmatrix} x \\ u \end{bmatrix}^{\top} \underbrace{\begin{bmatrix}
			q_{xx}^{*} & q_{xu}^{*} \\ q_{xu}^{*\top} & q_{uu}^{*}
	\end{bmatrix}}_{Q^{*}} \begin{bmatrix} x \\ u \end{bmatrix} + \frac{\gamma \trace(P\Sigma)}{1-\gamma}, 
\end{gather}
where $q_{xx}^{*} = L_{xx} + \gamma A^{\top}PA$, $q_{xu}^{*} = L_{xu} + \gamma A^{\top}PB$ and $q_{uu}^{*} = L_{uu} + \gamma B^{\top}PB$. 
The matrix $P \in \mathbb{R}^{n_{x}\times n_{x}}$ is the solution to the discrete-time algebraic Riccati equation
\begin{equation}\label{DARE}
	P = Q^{*}/q_{uu}^{*},
\end{equation}
and $Q^{*}/q_{uu}^{*} = q_{xx}^{*} - q_{xu}^{*}q_{uu}^{*-1}q_{xu}^{*\top}$ denotes the Schur complement of block $q_{uu}^{*}$ of matrix $Q^{*}$. Under Assumption \ref{assumtion stabilizability}, $P$ can be shown to be positive definite and unique \cite[Cond. 6.1.32]{DavisStochasticControl}. The optimal policy can then be found by minimizing \eqref{qstar} with respect to $u$, resulting in 
\begin{equation}\label{optimallinearpolicy}
	\pi^*(x) = -q_{uu}^{*-1}q_{xu}^{*\top}x.
\end{equation}
\begin{theorem}\label{Theorem fixed point}
	Under linear quadratic assumptions \eqref{linearmap}-\eqref{quadraticcost} and Assumption \ref{assumtion stabilizability}, the unique fixed point of $\hF$ is
	\begin{equation}\label{optimal q hat}
		\hat{q}(x,u) = q^{*}(x,u) + \Delta e,
	\end{equation} 
	where the constant $\Delta e = \frac{\gamma \emph{Tr}(q_{xu}^{*}q_{uu}^{*-1}q_{xu}^{*\top}\Sigma)}{1-\gamma} \ge 0$. The associated policy is then $ \hat{\pi}(x) = \arg \min_{u\in\mathbf{U}} \hat{q}(x,u) = \pi^*(x)$.
\end{theorem}
\begin{pf}
	We consider quadratic functions
	\begin{equation*}
		q(x,u) = \begin{bmatrix} x \\ u \end{bmatrix}^{\top} \underbrace{\begin{bmatrix}
				q_{xx} & q_{xu} \\ q_{xu}^{\top} & q_{uu}
		\end{bmatrix}}_{Q} \begin{bmatrix} x \\ u \end{bmatrix} + e,
	\end{equation*}
	with $Q$ positive definite, and seek to find a solution $\hat{q}$ to
	\begin{equation}\label{fixed point equation}
		q(x,u) = \hF q (x,u),
	\end{equation}
	within this class. Note that $q(x,u)\in \mathcal{S}(\mathbf{K})$ and the solution to \eqref{fixed point equation} is unique according to Proposition \ref{Prop monotonicity}. Then,
	\begin{equation}\label{ciao}
		\hF q(x,u) = \ell(x,u) + \gamma \inf_{w} (\#),
	\end{equation}
	where
	\begin{align*}
		(\#) & = \E_\xi \Bigg[ \begin{bmatrix}Ax+Bu+\xi \\ w \end{bmatrix}^{\top} Q
		\begin{bmatrix}Ax+Bu+\xi \\ w \end{bmatrix} + e\Bigg] \\
		& = \begin{bmatrix}Ax+Bu\\ w \end{bmatrix}^{\top} Q \begin{bmatrix}Ax+Bu\\ w \end{bmatrix} + \trace\left(q_{xx}\Sigma\right)+e.
	\end{align*}
	Since $Q$ is positive definite by assumption, we have that $(\#)$ is minimized by 
	\begin{equation}\label{minimizingw}
		w = -q_{uu}^{-1}q_{xu}^{\top}(Ax+Bu).
	\end{equation}
	If we substitute \eqref{minimizingw} into \eqref{ciao}, we can write \eqref{fixed point equation} as
	\begin{align*}
		& \begin{bmatrix} x \\ u \end{bmatrix}^{\top} Q \begin{bmatrix} x \\ u \end{bmatrix} + e = \gamma (\trace\left(q_{xx}\Sigma\right)+e) + \\
		& \quad\quad\quad + \begin{bmatrix} x \\ u \end{bmatrix}^{\top} \left( L + \gamma\begin{bmatrix} A^{\top} \\  B^{\top} \end{bmatrix}
		(Q/q_{uu})
		\begin{bmatrix} A & B \end{bmatrix} \right) \begin{bmatrix} x \\ u \end{bmatrix}.
	\end{align*}
	Since \eqref{fixed point equation} has to hold for all $(x,u)\in \mathbf{K}$, we impose
	\begin{subequations}{}\label{cases}
		\begin{align}
			Q & = L + \gamma \begin{bmatrix} A^{\top} &  B^{\top} \end{bmatrix}^{\top}
			(Q/q_{uu})
			\begin{bmatrix} A & B \end{bmatrix}, \label{cases1}\\
			e & = \gamma (\trace\left(q_{xx}\Sigma\right)+e). \label{cases2}
		\end{align}
	\end{subequations}
	Note that $Q^{*}$ defined in \eqref{qstar} is a solution to \eqref{cases1}. In fact, by exploiting \eqref{qstar}-\eqref{DARE}, we can decompose $Q^{*}$ as
	\begin{align*}
		Q^{*} & = L + \gamma \begin{bmatrix}
			A^{\top}PA & A^{\top}PB \\ B^{\top}P^{\top}A & B^{\top}PB \end{bmatrix} \\
		& = L + \gamma \begin{bmatrix} A^{\top} &  B^{\top} \end{bmatrix}^{\top}
		(Q^{*}/q_{uu}^{*})
		\begin{bmatrix} A & B \end{bmatrix}.
	\end{align*}
	Moreover, equation \eqref{cases2} is satisfied by 
	\begin{equation*}
		e = \frac{\gamma \trace(q_{xx}\Sigma)}{1-\gamma}.
	\end{equation*}
	Therefore, the unique solution to \eqref{fixed point equation} is
	\begin{align*}
		\hat{q}(x,u) & = \begin{bmatrix} x \\ u \end{bmatrix}^{\top} Q^{*} \begin{bmatrix} x \\ u \end{bmatrix} + \frac{\gamma \trace(q^{*}_{xx}\Sigma)}{1-\gamma} \\
		& = q^{*}(x,u) + \Delta e.
	\end{align*}
Finally, $\hat{\pi}(x) = \arg \min_{u\in\mathbf{U}} \hat{q}(x,u) = \pi^\ast(x)$. \qed 
\end{pf}
We can conclude that, in the linear quadratic (LQ) case, the optimal solution $\hat{q}$ to the RLP \eqref{alternativeLPQfunctions} preserves the shape of the optimal $q$-function; in particular, $\hat{q}$ and $q^*$ share the same minimizer with respect to $u$, hence give rise to the same optimal policy. We also note that, in accordance with Proposition \ref{Prop difference}, $q^* \le \hat{q}$. In case of deterministic LQ systems the RLP reduces to \eqref{LPdeterministic} and, since $\Sigma=0$, $\hat{q}=q^*$, confirming the results in \cite{GoranADP}. Finally, we would like to mention that linear affine systems, non-zero mean noise and general quadratic cost functions can in principle be treated in a similar way to the LQ problem above by allowing linear terms in the $q$-function \cite{BoydGenQuadratics}.

\section{Data-driven implementation}

In this last section, we aim to provide a comparison between the classical LP formulation \eqref{LPoldQfunctions} and the RLP \eqref{alternativeLPQfunctions} in terms of scalability and performance. We stress again that all the LPs introduced in this paper, in general, are not directly solvable. First, $q$ is an optimization variable in the infinite dimensional space $\mathcal{S}(\mathbf{K})$. As customary in the LP framework (see e.g. \cite{SchweitzerPolynomApproxinMDP,deFariasLPapproach}), we can restrict $q$ to the span of a finite family of basis functions. In the following experiments, we will consider $q$ to be confined to the finite dimensional subspace of quadratics
\begin{equation*}
	\tilde{\mathcal{S}}(\mathbf{K}) = \Bigg\{ q \in \mathcal{S}(\mathbf{K}) \;\; \bigg| \;\; q = \begin{bmatrix} x \\ u \end{bmatrix}^{\top} Q \begin{bmatrix} x \\ u \end{bmatrix} + e \Bigg\},                                                                          \end{equation*}                                                                                                                                               
where $Q=Q^{\top}\in\mathbb{R}^{(n_{x}+n_{u})\times (n_{x}+n_{u})}$ and $e\in\mathbb{R}$. A similar subspace $\tilde{\mathcal{S}}(\mathbf{X})$ of quadratics is used to replace the infinite dimensional space $\mathcal{S}(\mathbf{X})$ in \eqref{LPoldQfunctions}. Note that, under non-degeneracy assumptions on $Q$, this implies that the associated policies will be of the form $u=Kx$. For simplicity, we take $c$ to be a probability measure with zero mean and covariance matrix $C$. In this case, as pointed out in \cite{PaulADP}, we can compute the objectives in \eqref{LPoldQfunctions} and \eqref{alternativeLPQfunctions} without having to perform a high dimensional integration as
\begin{equation}\label{simplified objective function}
	\int_{\mathbf{K}} q(x,u)c(dx,du) = \E_cq = \trace(QC) + e.
\end{equation}
In general, if $q^* \in \tilde{\mathcal{S}}(\mathbf{K})$ ($v^* \in \tilde{\mathcal{S}}(\mathbf{X})$), then the choice of $c$ is not relevant. If, however, $q^* \notin \tilde{\mathcal{S}}(\mathbf{K})$, then the choice of $c$ can influence the quality of the solution. Moreover, in the latter case, the output of the LPs can significantly differ from the actual cost of playing the extracted policy \textit{a-posteriori}, usually called \textit{online performance}. Theoretical bounds on this quantity do exist for the LP approach, but they tend to be very conservative \cite{deFariasLPapproach,PaulADP}. In practice, the only means of comparison is often empirical observation of the resulting policies, an approach pursued in Experiment 3.

A second source of intractability is that the LPs have an infinite number of constraints. To tackle this, we replace the infinite constraints with a finite subset, as argued in \cite{deFariasConstraintSampling,AlexandrosIFAC20,GoranADP}, by letting the system interact with the environment and collecting roll-out data $(x,u,x^{+}_u, \ell(x,u))$. In the numerical experiments we generate roll-outs by sampling $(x,u)$ pairs according to given distributions. For each pair we assume we can measure the corresponding $\ell(x,u)$ and estimate the expectation over $x_u^+$ that appears in the constraint of the LPs by Monte Carlo sampling. We note that once the roll-outs are available one no longer needs the model (the functional form of $f$ and $\ell$) to formulate the LP. In the spirit of \cite{GoranADP,AlexandrosIFAC20}, one can think of the resulting LP that approximates $q^*$ as solving an RL problem, where the goal is to learn a policy without knowing the dynamics and cost function.

In the numerical experiments reported below we work in unconstrained spaces $\mathbf{X} = \mathbb{R}^{n_{x}}$, $\mathbf{U} = \mathbb{R}^{n_{u}}$ and $\mathbf{\Xi} = \mathbb{R}^{n_{\xi}}$. In Tab. \ref{table1} we report the discount factor, stage cost, covariance matrix of $c$, distributions used to sample states $x$ and inputs $u$, noise distribution, number of constraints employed and number of samples (MC) used in the Monte Carlo approximation of the expectation. 

\begin{table*}
	\centering \scriptsize
	\begin{threeparttable}
	\caption{Parameters employed in the three experiments. $\mathcal{U}$, $\mathcal{N}$, diag and $I_{n}$ denote the uniform and normal distribution, a block diagonal matrix and the $n\times n$ identity matrix, respectively. Simulations are run on a processor Intel Core i7-7560U, implemented in Python using Pytorch to build the LPs and with GUROBI as solver. Code available at \url{https://github.com/gargiani/ADP_LP}.}
	\label{table1}
	\begin{tabular}{lllllllll} 
		\toprule
		 & $\gamma$ & $L$ & $C$ & State distr. & Input distr. & Noise distr.& \#constr. & MC \\ 
		\midrule
		\textbf{1} & 0.95 & diag($I_2,10^{-2}$) & diag($I_2,10^{-1}$) & $\mathcal{U}([-3,3]^2)$ & $\mathcal{U}([-1,1])$ & $\mathcal{N}(0,10^{-6}I_2)$ & $\le 2\cdot10^4$ & $10^2$  \\
		\textbf{2} & 0.95 & diag($I_{n_x},10^{-4}$) & diag($I_{n_x},0.8I_2$) & $\mathcal{U}\left([-0.5, 0.5]^{n_x} \right)$ & $\mathcal{U}\left([-3, 3]^2\right)$ & $\mathcal{N}(0,10^{-4}I_{n_x})$& $5\cdot10^4$ & $10^2$ \\ 
		\textbf{3} & 0.99 & $\textrm{diag}(I_2,10^2, 10,10^{-3})$ & diag($I_4,0.8$) & $\mathcal{U}\left([-3, 3]^2\times[-1, 1]^2 \right)$ & $\mathcal{U}\left([-100, 100] \right)$ & $\mathcal{N}(0,10^{-6}I_4)$ & $10^4$ & 1 \\  
		\bottomrule
	\end{tabular}
	\end{threeparttable}
\end{table*}

\begin{figure}
	\centering
	\begin{subfigure}[t]{0.49\linewidth}
		\includegraphics[scale=0.28]{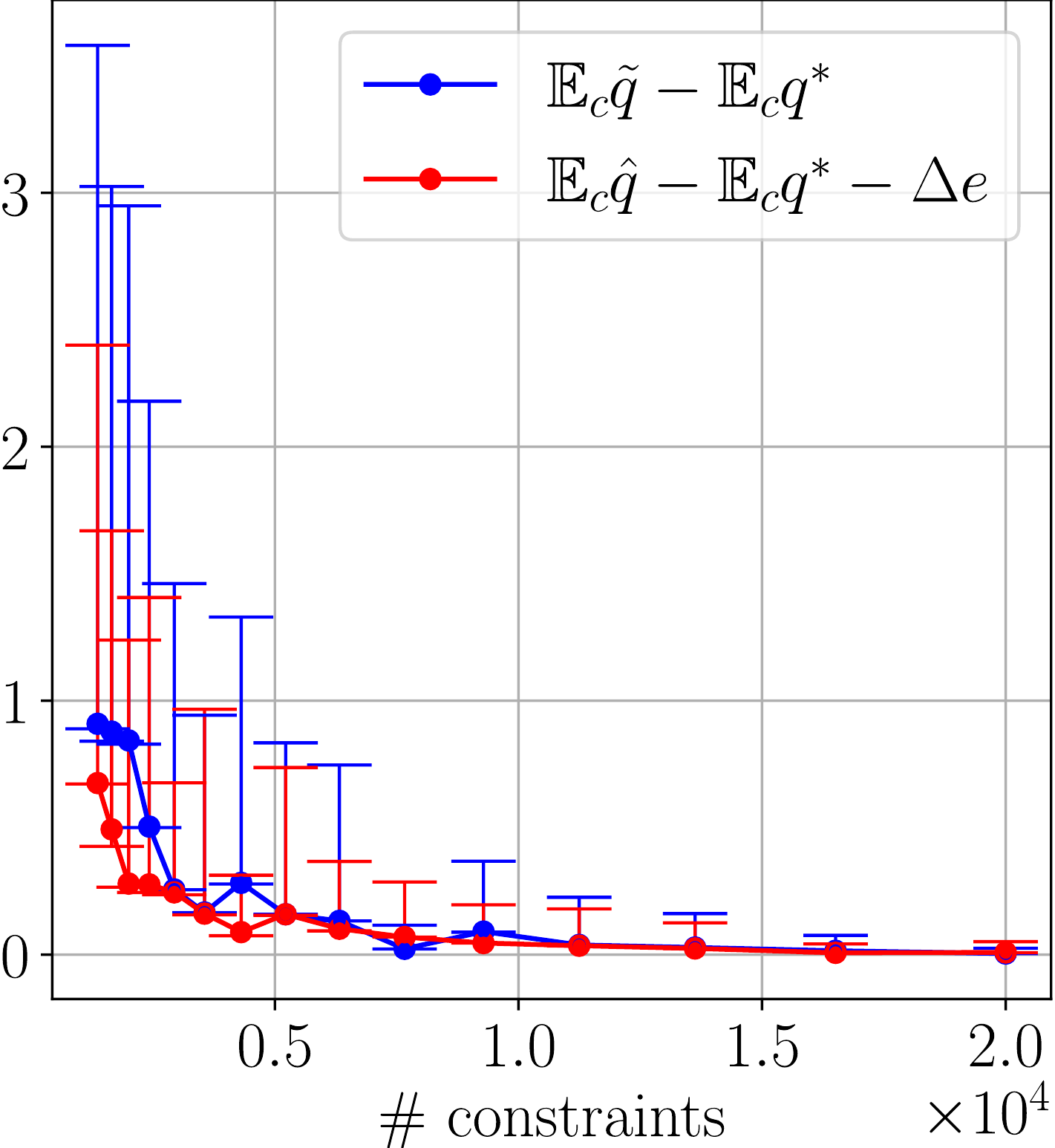}
		\caption{\scriptsize Performance vs constraints}
		\label{fig:linear_performance_constraints}
	\end{subfigure}%
	\begin{subfigure}[t]{0.505\linewidth}
		\includegraphics[scale=0.278]{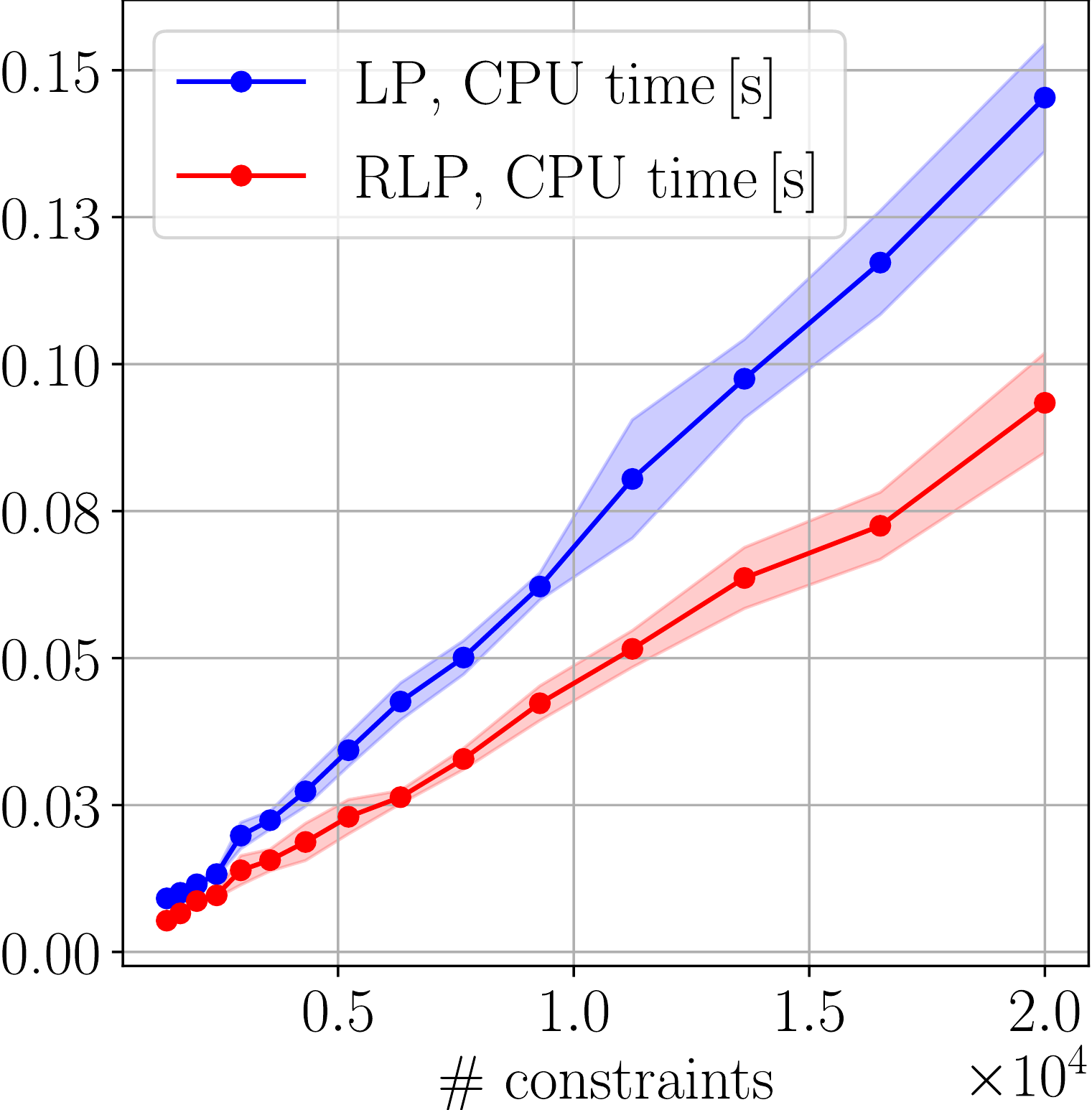}
		\caption{\scriptsize Time vs constraints}
		\label{fig:linear_time_constraints}
	\end{subfigure}
	\par\bigskip 
	\begin{subfigure}[t]{0.5\linewidth}
		\includegraphics[scale=0.282]{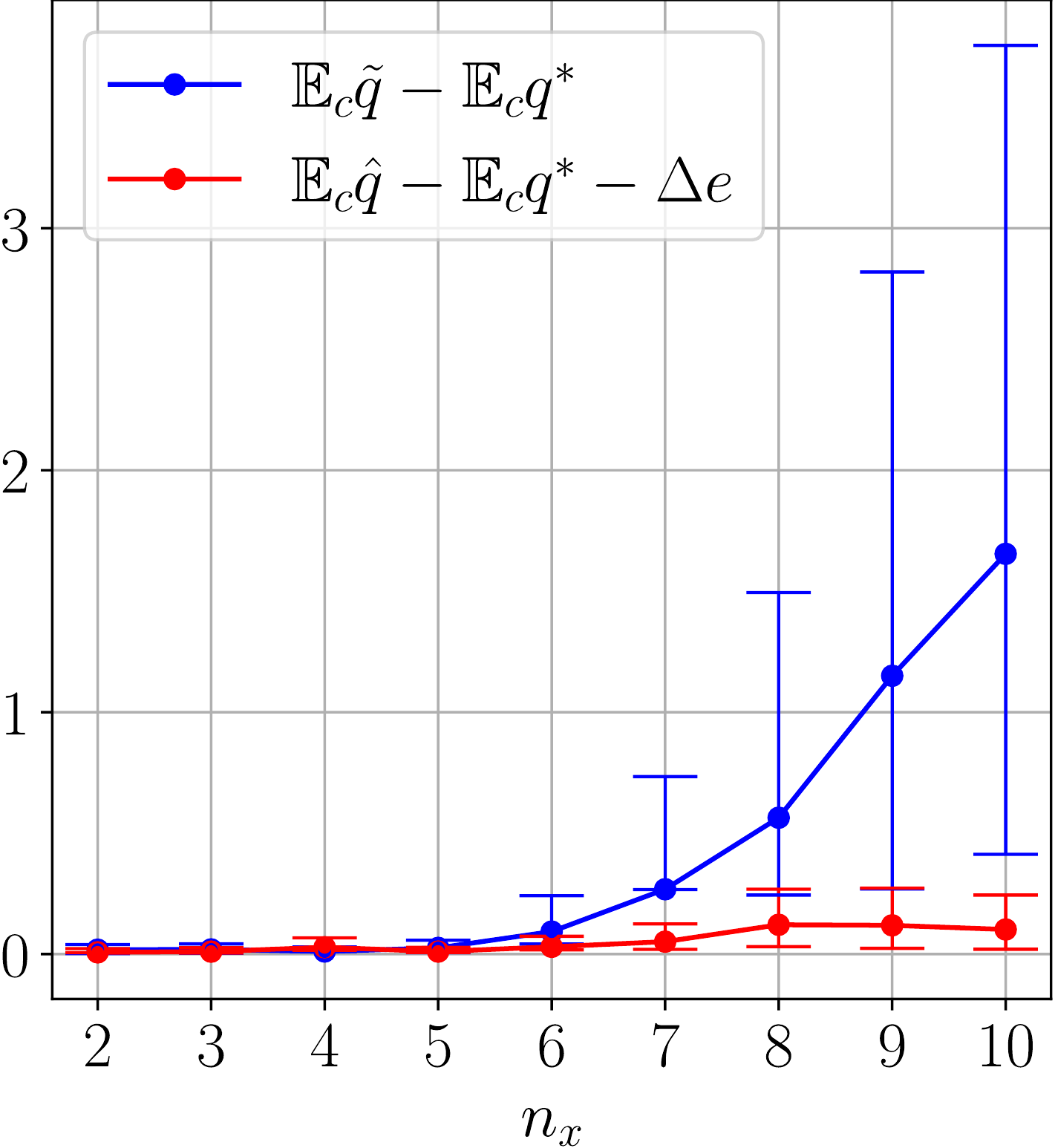}
		\caption{\scriptsize Performance vs dimension}
		\label{fig:linear_performance_dimension}
	\end{subfigure}%
	\begin{subfigure}[t]{0.5\linewidth}
		\includegraphics[scale=0.286]{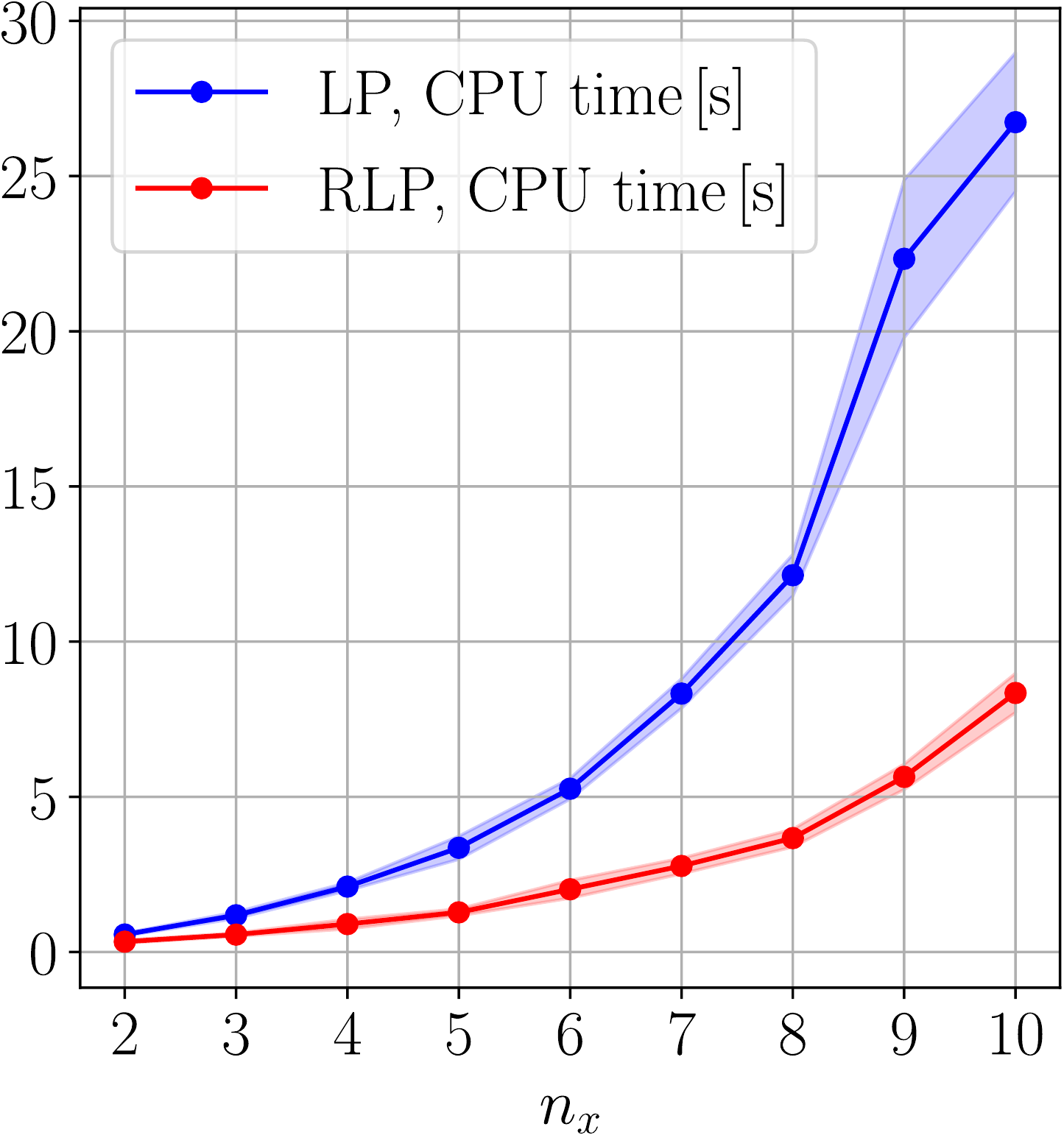}
		\caption{\scriptsize Time vs dimension}
		\label{fig:linear_time_dimension}
	\end{subfigure}
	\caption{Optimality gap and CPU time as a function of the number of constraints and state dimension for the LP and RLP. We show the realization of 10 identical runs, depicting the realization amplitude with vertical bars and the standard deviation with solid bands around the mean value. Finally, note that in Figs. \ref{fig:linear_performance_constraints} and \ref{fig:linear_performance_dimension} the RLP is debiased from the constant up-shift (see Theorem \ref{Theorem fixed point}) for comparison purposes.}
	\label{linear_dimensionality}
\end{figure}

\textbf{Experiment 1: linear systems and constraints.} Consider the problem of learning an optimal control policy for the following unknown LTI system
\begin{equation*}\label{system experiment}
	x_{k+1} = \begin{bmatrix}
			1 & 0.1 \\ 0.5 & -0.5
	\end{bmatrix}x_{k} + \begin{bmatrix}
			1 \\ 0.5
	\end{bmatrix}u_{k} + \xi_{k}.
\end{equation*}
Fig. \ref{fig:linear_performance_constraints} shows that both the solutions to the LP ($\tilde{q}$) and the RLP ($\hat{q}$) converge to the optimal $q$-function as we increase the number of sampled constraints, though the RLP displays slightly smaller variability. We observe in Fig. \ref{fig:linear_time_constraints} that the solving time of both programs grows linearly with the number of constraints but, due to fewer decision variables, the RLP has a smaller slope.

\textbf{Experiment 2: linear systems and dimension.} We randomly generate LTI systems with increasing number of states and 2 inputs
\begin{equation*}
	x_{k+1} = A x_k + B u_k + \xi_k,
\end{equation*}
where $A\in \mathbb{R}^{n_x\times n_x}$, $B\in \mathbb{R}^{n_x\times 2}$, $n_{x} = \{ 2,\cdots,10 \}$, $A_{ii} = 0.5$, and $A_{ij}$ and $B_{ij}$ with $j\neq i$ generated in an Erd\H{o}s-R\'{e}nyi  fashion according to
\[  A_{ij},B_{ij} \sim 
\begin{cases}
	0 & \text{with probability } 0.1\\
	\mathcal{U}([-0.1, 0.1]) & \text{with probability }0.9,
\end{cases} \]
ensuring that Assumption \ref{assumtion stabilizability} is satisfied. Figs. \ref{fig:linear_performance_dimension}--\ref{fig:linear_time_dimension} show how the simpler structure of the RLP results in both better performance and greater scalability. In fact, given a fixed number of constraints, the same for both programs, the RLP substantially outperforms the LP on the optimality gap, solving time and solution variability as the order of the system increases. 

%
%

\textbf{Experiment 3: nonlinear system.} Finally, we consider an inverted pendulum on a cart
\begin{equation*}
	\begin{aligned}
		\ddot{p} & =\frac{1}{(M+m)}\left(  m\ell \ddot{\theta}\cos{\theta} - m\ell\dot{\theta}^2\sin{\theta} + u\right),\\
		\ddot{\theta} & = \frac{1}{\ell}\left( g\sin{\theta} + \ddot{p}\cos{\theta}\right), 
	\end{aligned}
\end{equation*}
with $M = 4\,\,\textrm{kg}$, $m = 2\,\,\textrm{kg}$, $\ell = 1\,\,\textrm{m}$, $g = 9.8\,\,\textrm{m}/\textrm{s}^2 $, state $x = (p,\dot{p},\theta,\dot{\theta})$                                                                                          comprising position and linear velocity of the cart, angular position and velocity of the pole, and input force $u$ acting on the cart. The system is discretized with forward Euler (sampling time $10^{-3}$ s) and perturbed with additive disturbance according to Tab.~\ref{table1}. 

\begin{figure}
	\centering
	\begin{subfigure}{0.5\linewidth}
		\includegraphics[width=\linewidth]{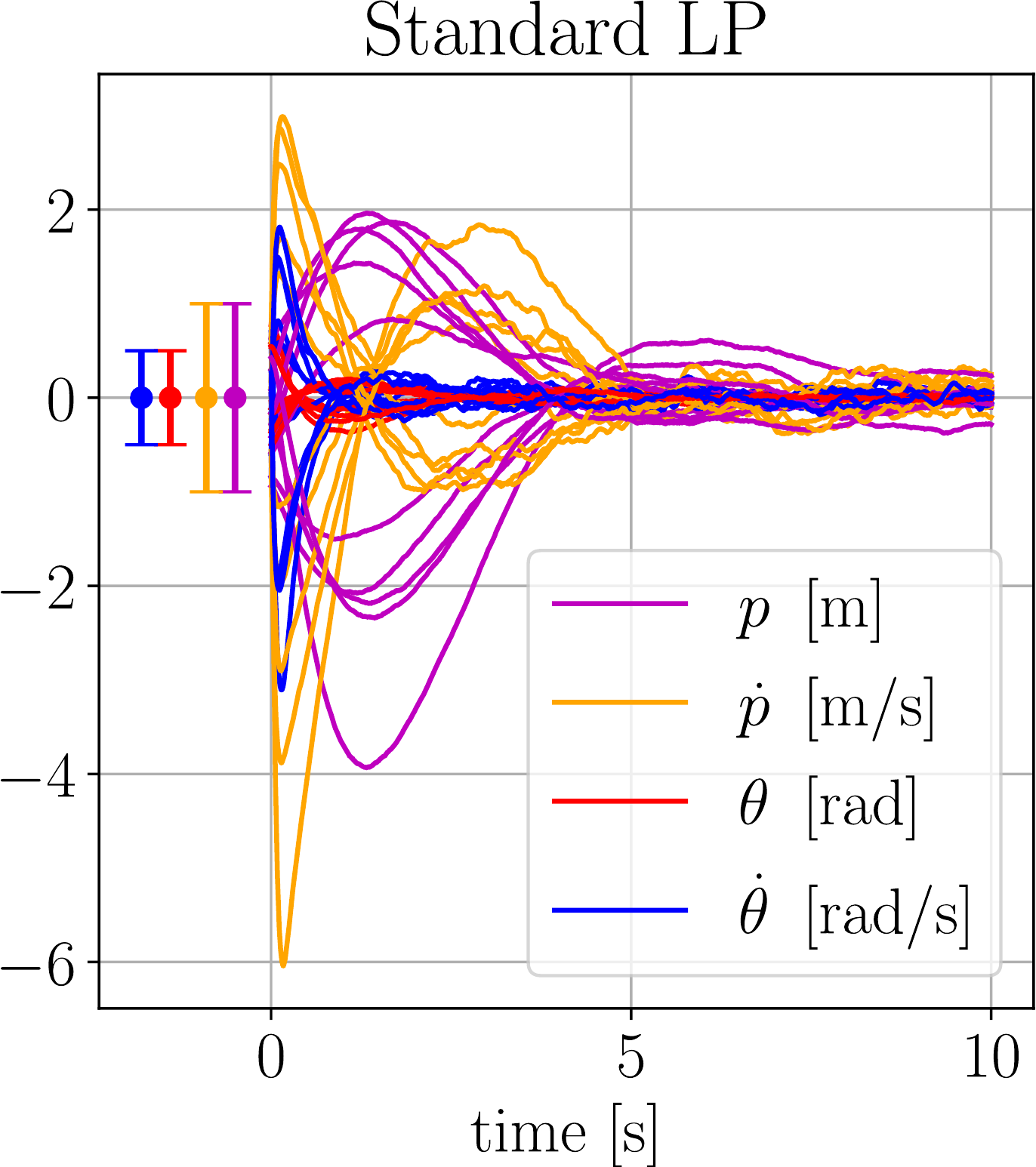}
	\end{subfigure}%
	\begin{subfigure}{0.5\linewidth}
		\includegraphics[width=\linewidth]{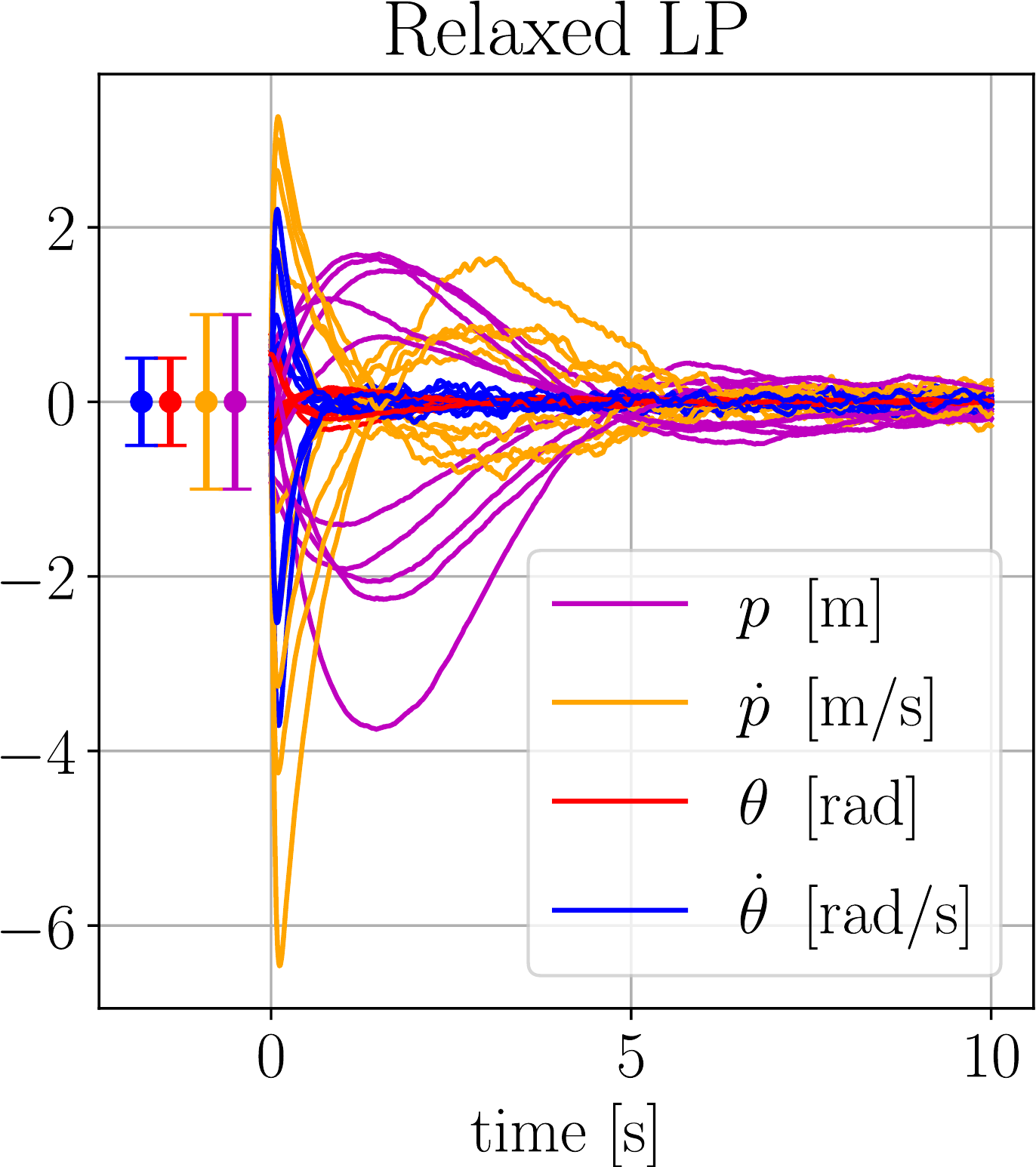}
	\end{subfigure}
	\caption{Trajectories of the cart-pole system controlled with the policies derived from the LP (left) and RLP (right). The vertical bars denote the initial state distribution $\nu_0$.}
	\label{cart-pole}
\end{figure}

Tab.~\ref{table2} shows that the RLP solving time is still faster. This is expected as once the subspaces $\tilde{\mathcal{S}}(\mathbf{K})$ and $\tilde{\mathcal{S}}(\mathbf{X})$ are fixed the structure of the linear programs do not depend on whether the dynamics is linear or nonlinear. The resulting nonlinear closed-loop trajectories are depicted in Fig \ref{cart-pole}, where the cart-pole dynamics is initialized 10 times for each method with initial state distribution $\nu_0$ given by $x_{\nu_0} \sim \mathcal{U}([-1, 1]^2 \times [-0.5, 0.5]^2)$. Both policies succeed in stabilizing the system around the vertical position. Moreover,  Tab.~\ref{table2} shows that the discounted costs incurred by playing the LP and RLP policies on the nonlinear system are similar, and in both cases outperform the (model-based) linear quadratic regulator (LQR) computed for the system linearized about $x=0$. In general, however, the relative performance between the LP and RLP non-trivially depends on the specific parametrization of the problem, hence, on the geometry of the linear program. A similar consideration was discussed in \cite{DesaiADPSmoothedLP} after adding a slack variable in the constraints of the standard LP.

\begin{table}
	\centering \scriptsize 
	\begin{threeparttable}
	\caption{Costs incurred by three different policies on the cart-pole system and corresponding CPU times.}
	\label{table2}
	\begin{tabular}{llll} 
		\toprule
		& LP & RLP & LQR \\ 
		\midrule
		$\mathbb{E}_{\nu_0} v$ & $2.92\cdot10^4$ & $2.91\cdot10^4$ &  3.98$\cdot10^4$\\
		CPU time [s] & 0.884 & 0.546 & $-$ \\
		\bottomrule
	\end{tabular}
	\end{threeparttable}
\end{table}


\section{Conclusions}

We introduced a relaxed Bellman operator $\hF$ for $q$-functions, we proved that it retains the key properties of monotonicity and contractivity with respect the the sup-norm, and we built a RLP that retrieves the fixed point of $\hF$. We characterized this fixed point in the LQ setting and showed that it preserves the minimizer of the optimal function $q^*$, hence recovers the optimal policy. The RLP provides significant improvements with respect to the classical LP formulation in terms of convergence to the optimal policy and scalability. 

A major difference with respect to the classical LP approach is that we are not approximating the optimal $q$-function directly, but we look for functions that sufficiently preserve the minimizer of $q^*$ while making the linear program more efficient. Current and future research directions also include relaxing the LQ assumptions by enlarging the function classes of $f$ and $\ell$ and characterizing again the fixed point. Convergence comparison with other model-free ADP methods is another valuable direction. Moreover, significant insights about the exploration logic are needed for the LP approaches in general. In fact, the approximation quality is non-trivially affected by the sampling design, and a poor choice of parameters can often lead to bad performance or unboundedness issues. Finally, tighter error bounds could be provided to link solution quality with online performance.

\bibliographystyle{plain}        
\bibliography{autosamfile}           



\end{document}